\documentclass[a4paper]{iopart}
\usepackage{graphicx}
\usepackage{dcolumn}
\usepackage{bm}
\usepackage{xcolor}
\usepackage{qcircuit}
\usepackage{braket}
\usepackage{float}
\usepackage{algorithm}
\usepackage{algorithmicx}
\usepackage{algpseudocode}
\usepackage{hyperref}

\renewcommand{\figureautorefname}{Figure~\negthinspace}

\begin{document}


\title{An end-to-end trainable hybrid classical-quantum classifier}

\author{Samuel Yen-Chi Chen}
\ead{ychen@bnl.gov}
\address{%
 Computational Science Initiative, Brookhaven National Laboratory
}%
\author{Chih-Min Huang}
\ead{b06501134@ntu.edu.tw}
\address{%
 Department of Physics, National Taiwan University, Taipei 10617, Taiwan}%
\author{Chia-Wei Hsing}
\ead{cwhsing0219@gmail.com}
\address{%
 Department of Physics, National Taiwan University, Taipei 10617, Taiwan}%
\author{Ying-Jer Kao}
\ead{yjkao@phys.ntu.edu.tw}
\address{%
 Department of Physics, National Taiwan University, Taipei 10617, Taiwan}%

\date{\today}

\begin{abstract}
%
We introduce a hybrid model combining  a quantum-inspired tensor network and  a variational quantum circuit  to perform supervised learning tasks.
This architecture allows for the classical and quantum parts of the model to be  trained simultaneously, providing an end-to-end training framework. 
We show that compared to the principal component analysis,  a tensor network based on the matrix product state with low bond dimensions performs better as a feature extractor for the input data of the variational quantum circuit in the binary and ternary classification of MNIST and Fashion-MNIST datasets.
The architecture is highly adaptable and the classical-quantum boundary can be adjusted according the availability of the quantum resource by exploiting the correspondence between tensor networks and quantum circuits.

\end{abstract}

\submitto{Machine Learning: Science and Technology
}
\maketitle


\section{\label{sec:Indroduction}Introduction}
Quantum computing (QC) has demonstrated superiority in problems  intractable on classical computers \cite{harrow2017quantum, nielsen2002quantum}, such as factorization of large numbers \cite{shor1999polynomial} and  search in an unstructured database \cite{grover1997quantum}. 
Recent growth of the quantum volume in noisy intermediate-scale quantum (NISQ) \cite{preskill2018quantum} devices has stimulated rapid development in circuit-based quantum algorithms. 
Due to the  noise associated with the quantum gates and lack of quantum error correction on NISQ devices, performing quantum computation with large circuit depth is impossible currently. 
It is therefore highly desirable to develop quantum algorithms that are resilient to noise with moderate circuit depth. 
{Variational quantum algorithms}  \cite{cerezo2020variational} are a class of algorithms currently under rapid development in many fields.   
In particular, quantum machine learning (QML)~\cite{schuld2018supervised,biamonte2017quantum,dunjko2018machine} using variational quantum circuits (VQC) shows great potential in surpassing the performance of classical machine learning (ML).
%
%
One of the major advantages of VQC-based QML compared to its classical counterpart is the drastic reduction in the number of model parameters, potentially mitigating the problem of overfitting common in the classical ML. 
Moreover, it has been shown that under certain conditions, QML models may learn faster or achieve higher testing accuracies than its classical counterpart~\cite{chen2020quantum,chen2020qcnn}.
A modern QML architecture typically includes a classical  and a quantum part.
Famous examples in this hybrid genre include quantum approximate optimization algorithm~\cite{farhi2014quantum}, and quantum circuit learning~\cite{mitarai2018quantum}, where the VQC plays an crucial role as the quantum component with the circuit parameters  updated via a classical computer.
Various architectures and geometries of VQC have been suggested for tasks ranging from binary classification \cite{chen2020qcnn,abohashima2020classification,schuld2018circuit,mitarai2018quantum} to reinforcement learning \cite{chen19,lockwood2020reinforcement,wu2020quantum}.


One of the key challenges in the NISQ era is that available quantum hardwares have   limited quantum  volume and are only capable of executing quantum operations with small circuit-depth. 
That means that most of the dataset commonly used for classical ML tasks are too large for the NISQ devices. 
To process the data with input dimension exceeding the number of available qubits, it is necessary to apply  dimensional reduction techniques to first compress the input data. 
For example, in Ref.~\cite{mari2019transfer},  pre-trained classical deep convolutional neural network is used to compress the high-resolution images into a low dimension representation.
However, since the pre-trained model there has a huge number of parameters, it is not clear what is the contribution of the quantum circuit in the whole workload. 
On the other hand, a major challenge in building a QML model is how to encode high-dimensional classical data into a quantum circuit efficiently. 
With the limitation imposed by NISQ in mind, the encoding process should be designed to consume as few gate operations as possible.
\emph{Amplitude encoding} is one of the encoding method which can provide significant advantage in terms of the number of qubits required to handle the input data. 
For an $N$-dimensional vector, amplitude encoding requires only $\log_2 N$ qubits; however, the quantum circuit depth to prepare such encoded state exceeds the current limits of NISQ devices. 
Other approaches like single-qubit rotations require only a shallow circuit but it is unclear how to employ such encoding schemes to load high-dimensional data into a quantum circuit.
This can be potentially mitigated by preprocessing the input data with classical methods to perform dimension reduction.
Principal component analysis (PCA) is a simple dimension reduction method and has been widely used in the QML research; yet it lacks the representation power to retain enough information. 
More powerful and expressive models such as neural networks are not commonly utilized due to the requirement of  pre-training and the significant number of parameters involved. 
Therefore, it is necessary to devise a data compression scheme which can be naturally integrated with VQC. 

In this paper, we propose a hybrid framework where  a matrix product state (MPS)~\cite{Ostlund:1995iz,Schollwock:2011kt}, the simplest  form of tensor networks (TN)~\cite{Orus:2014um},  is used as a feature extractor to produce a low dimensional feature vector.
This information is subsequently  fed into a VQC for classification. 
Unlike other QML schemes where the classical neural network  has to be pre-trained, our framework is trained as a whole. 
This end-to-end training indicates the quantum-classical boundary can be adjusted based on the available quantum resource.
Furthermore, since a MPS can be realized precisely by a quantum circuit~\cite{Huggins:2019kh},  it is possible to replace the classical component with a quantum circuit, making the scheme highly  adaptable.
Our scheme has shown to be superior in the  binary classification task for the MNIST dataset~\cite{Chen:2020zv}. 
Here, we apply the scheme to more difficult tasks such as the ternary classification of MNIST and the classification tasks of Fashion-MNIST. 

The rest of the paper is organized as follows. 
Section~\ref{sec:TensorNetwork} gives a brief introduction to tensor networks and their application in classical ML. 
Section~\ref{sec:VariationalQuantumCircuits} describes the VQC used in this study. 
Section~\ref{sec:HybridTNVQCArchitecture} introduces the hybrid TN-VQC architecture. 
The performance of the model is shown in Sec.~\ref{sec:ExpAndResults}.
Finally we conclude in Sec.~\ref{sec:Conclusion}.
\section{\label{sec:TensorNetwork}Tensor Network}
Tensor networks are efficient representation of data residing in high-dimensional space. 
Originally developed in the context of condensed matter physics, TNs  have gained  attention in the deep  learning community, for both theoretical understanding and computationally efficiency~\cite{Orus:2019nr}.  
They have provided new inspiration for machine learning algorithms and showed encouraging success in both discriminative~\cite{Levine:2018qp,Stoudenmire:2018wk,Liu:2019ty,Reyes:2020fd} and generative learning tasks~\cite{Han:2018rt}.
In addition, the quantum entanglement inherent in the formulation of  tensor networks  points to a new direction in understanding the mechanism of deep neural networks and may provide a better way to design new network architectures~\cite{Levine:2019xt,Levine:2018qp}.

\begin{figure}[tbp]
\center
\includegraphics[width=.6\linewidth]{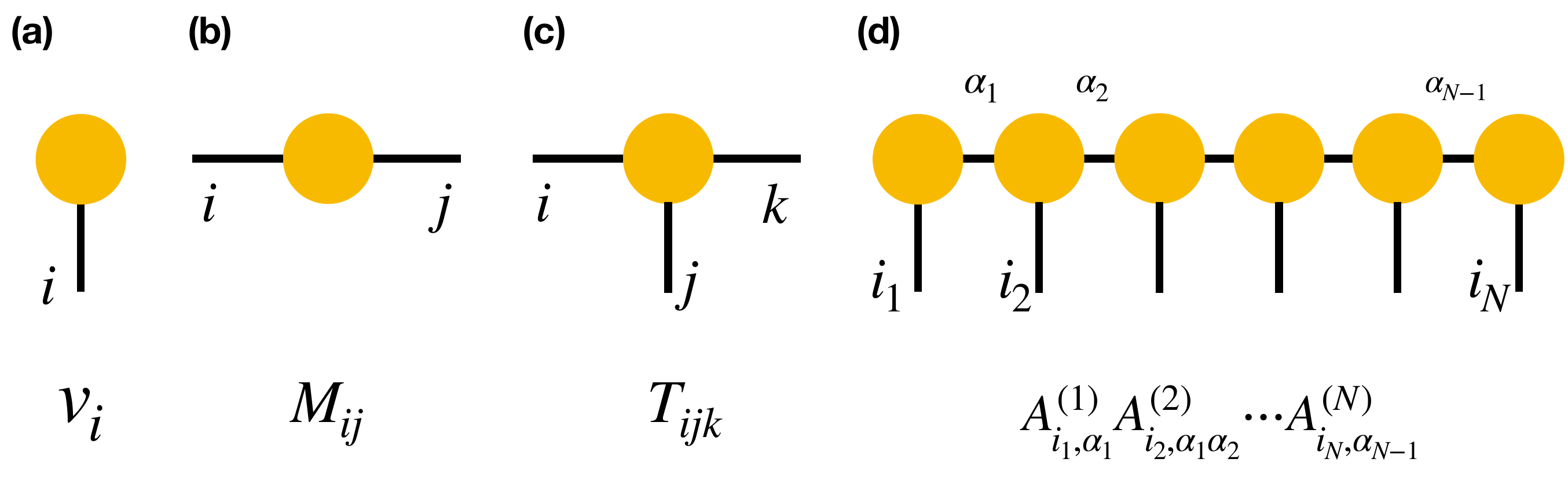}
\caption[Graphical Notation for Tensors and Tensor Networks]{{\bfseries Graphical Notation for Tensors and Tensor Networks.} Graphical tensor notation for (a) a vector, (b) a matrix,  (c) a rank-3 tensor and (d) a MPS. Here we follow the Einstein convention that repeated indices, represented by internal lines in the diagram, are summed over.
 }
\label{fig:TN}
\end{figure}

It is common to use graphical notation to express tensor networks. 
A tensor is represented as a closed shape, typically a circle, with emanating lines representing tensor indices (Fig.~\ref{fig:TN}).
The joined line indicates the corresponding index is contracted, as in the Einstein convention where repeated indices are summed over. 
The MPS, also known as tensor train,  is the simplest TN, and  the most widely used tensor networks in physics to study low-dimensional quantum systems, and has recently found application in the field of  machine learning~\cite{Cohen:2016mi,Stoudenmire:2016ve,Bengua:2015qf,Novikov:2015kq,Liu:2019ty,Efthymiou:2019qc}
In an MPS,  tensors are contracted through the ``virtual'' indices ($\alpha$'s in Fig.~\ref{fig:TN}(d)). 
The dimension of these virtual indices are called bond dimension and is indicated by $\chi$.
In the MPS representation of a quantum wave function, the bond dimension indicates the amount of quantum entanglement the MPS can represent in the bond. 
In the context of ML, this corresponds to the representation power of the MPS. 
%
%

In the current study, we choose the MPS as our TN for simplicity; there are other examples of TN with  distinct entanglement structures such as  the tree tensor network (TTN), multi-scale entanglement renormalization ansatz (MERA) and  projected entangled pair state (PEPS). 
The successful application of a specific TN can also give insights into the hidden correlations in the data. 
The quantumness inherent in the TN gives it great advantage over other architectures in the application of QML.
In particular, since each TN can  be mapped to a quantum circuit,  it means that although in the current scheme, the TN is treated classically, it is possible to replace the whole or part of the TN component by an equivalent quantum circuit when more qubits are available.
This gives the current scheme the flexibility to move the quantum-classical boundary based on the available resources.

\begin{figure}[tbp]
\center
\includegraphics[width=.4\linewidth]{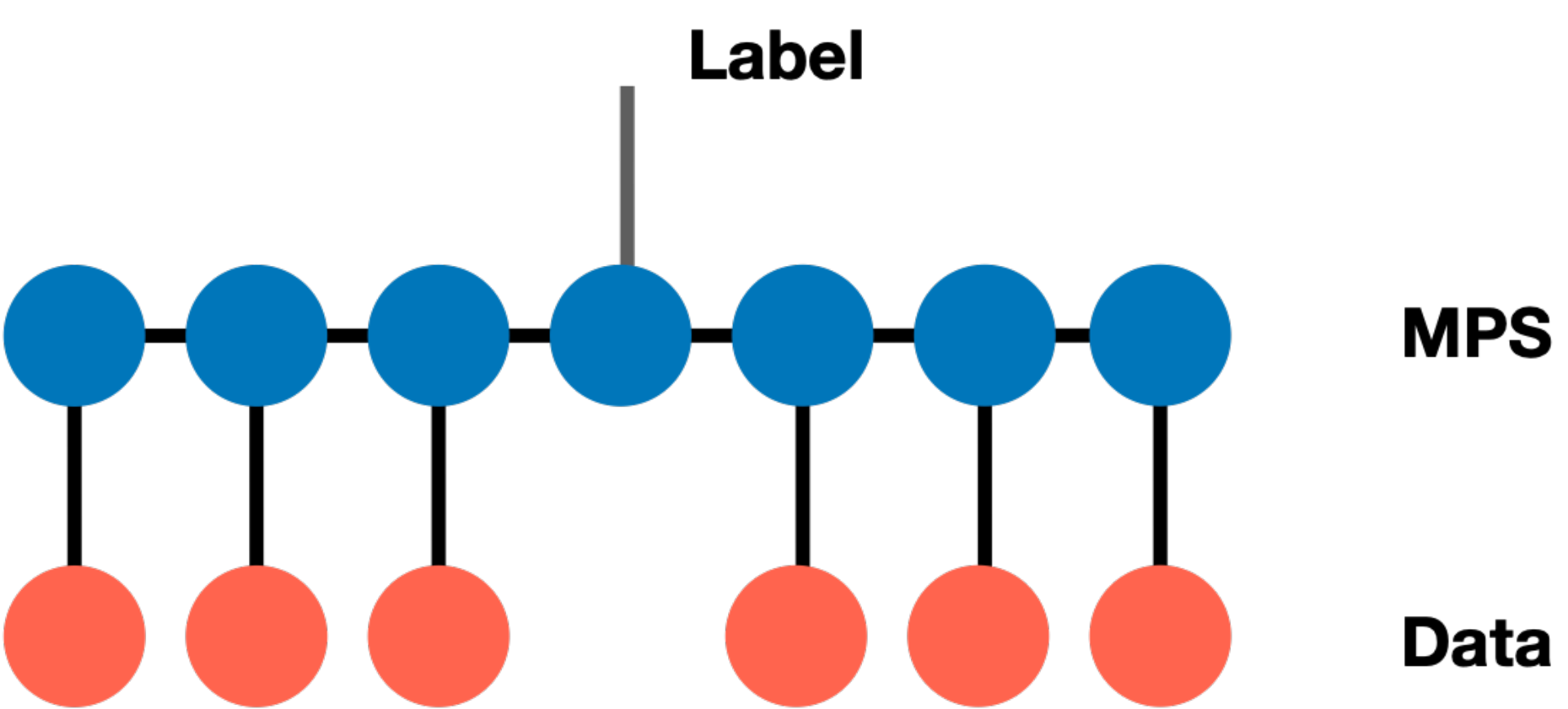}
\caption[MPS as a feature extractor.]{{\bfseries MPS as a feature extractor.} Data is encoded into a product state (blue nodes) which is contract with an MPS (yellow node) and a class label is generated. }
\label{fig:feature_MPS}
\end{figure}

We will use the MPS as a feature extractor to compress the input data. 
Following Ref.~\cite{Stoudenmire:2016ve}, we approximate a feature extractor  by the MPS decomposition as 
\begin{equation}
T_{i_{1} i_{2} \cdots i_{N}}^l=\sum_{\{\alpha\}} A_{i_{1}, \alpha_{1}}^{(1)} A_{i_{2}, \alpha_{1} \alpha_{2}}^{(2)} A_{i_{3}, \alpha_{2} \alpha_{3}}^{(3)}\cdots A_{i_j, \alpha_{j-1}\alpha_{j+1}}^{(j),l} \cdots A_{i_{N}, \alpha_{N-1}}^{(N)},
\end{equation}
%
 illustrated in Fig.~\ref{fig:feature_MPS}. 
 
%
%

\section{\label{sec:VariationalQuantumCircuits}Variational Quantum Circuit}
%
Variational quantum circuits originates from a quantum algorithm called \emph{variational quantum eigensolver}~\cite{peruzzo2014variational}, which is originally used to compute ground states. 
This family of algorithms have recently drawn significant attention and numerous efforts have been made to extend their applications~\cite{cerezo2020variational}. 
%
%
%
VQCs have been successfully applied to function approximation~\cite{mitarai2018quantum, chen2020quantum}, classification~\cite{mitarai2018quantum, schuld2018circuit,havlivcek2019supervised,Farhi2018ClassificationProcessors,benedetti2019parameterized, mari2019transfer, abohashima2020classification, easom2020towards, sarma2019quantum,chen2020qcnn}, generative modeling~\cite{dallaire2018quantum,stein2020qugan, zoufal2019quantum, situ2018quantum, nakaji2020quantum}, metric learning~\cite{lloyd2020quantum, nghiem2020unified}, deep reinforcement learning~\cite{chen19, lockwood2020reinforcement, jerbi2019quantum} , sequential learning \cite{chen2020quantum, bausch2020recurrent}, speech recognition \cite{yang2020decentralizing} and transfer learning~\cite{mari2019transfer}. 
It has been shown that VQCs are more expressive than conventional neural networks~\cite{sim2019expressibility,lanting2014entanglement,du2018expressive, abbas2020power} with respect to the number of parameters or the learning speed. 
It is demonstrated that with similar number of parameters, VQC-based models outperform classical models on testing accuracies~\cite{chen2020qcnn}, and achive optimal accuracy in function approximation tasks with fewer training epochs than their classical counterparts~\cite{chen2020quantum}.
Of particular interests for NISQ applications, it has been shown that such circuits are potentially resilient to noises in quantum hardware~\cite{kandala2017hardware,farhi2014quantum,mcclean2016theory}, and such robustness has been demonstrated empirically  on either noisy simulators or real quantum hardware~\cite{nghiem2020unified, chen19}. 
This strongly suggests that VQC-based architectures are suitable for building ML applications on NISQ devices. 

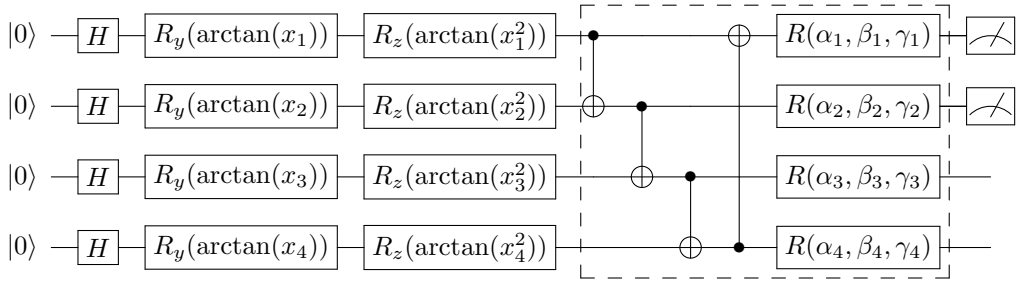
\begin{figure}
\begin{center}
\begin{minipage}{10cm}
\Qcircuit @C=1em @R=1em {
\lstick{\ket{0}} & \gate{H} & \gate{R_y(\arctan(x_1))} & \gate{R_z(\arctan(x_1^2))} & \ctrl{1}   & \qw       & \qw      & \targ    & \gate{R(\alpha_1, \beta_1, \gamma_1)} & \meter \qw \\
\lstick{\ket{0}} & \gate{H} & \gate{R_y(\arctan(x_2))} & \gate{R_z(\arctan(x_2^2))} & \targ      & \ctrl{1}  & \qw      & \qw      & \gate{R(\alpha_2, \beta_2, \gamma_2)} & \meter \qw \\
\lstick{\ket{0}} & \gate{H} & \gate{R_y(\arctan(x_3))} & \gate{R_z(\arctan(x_3^2))} & \qw        & \targ     & \ctrl{1} & \qw      & \gate{R(\alpha_3, \beta_3, \gamma_3)} &  \qw \\
\lstick{\ket{0}} & \gate{H} & \gate{R_y(\arctan(x_4))} & \gate{R_z(\arctan(x_4^2))} & \qw        & \qw       & \targ    & \ctrl{-3} & \gate{R(\alpha_4, \beta_4, \gamma_4)} & \qw \gategroup{1}{5}{4}{9}{.7em}{--}\qw 
}
\end{minipage}
\end{center}
\caption[Generic circuit architecture for the variational quantum classifier.]{{\bfseries Generic variational quantum circuit architecture.}
  The VQC component used in this work consists of  three parts:  the {encoding} part,  the {variational} part with parameters to be learned and  the {measurement} part which will output the Pauli-$Z$ expectation values via multiple run of the quantum circuit.  The quantum measurement would be performed on first $k$ qubits where $k$ is the number of classes. }

\label{Fig:Basic_VQC_Hadamard}
\end{figure}

 The VQC used in this work consists of three parts  (Fig.~\ref{Fig:Basic_VQC_Hadamard}): The first part is the \emph{encoding} part which consists of Hadamard gate $H$ and single qubit rotation gates $R_y(\arctan(x_i))$ and $R_z(\arctan(x_i^2))$, representing rotations along $y$-axis and $z$-axis by the given angle $\arctan(x_i)$ and $\arctan(x_i^2)$, respectively. 
The Hadamard gate $H$ is is used to create unbiased initial state as described \ref{encoding}. 
Notice the rotation angles $\arctan(x_i)$ and $\arctan(x_i^2)$ are for state preparation and  come directly from the input classical data. 
The data encoding part should be designed with respect to the problem of interest and plays a crucial role in the overall architecture~\cite{Schuld2018InformationEncoding}. 
Potential quantum advantage depends heavily  on the encoding scheme together with the hardware  limitations  incorporated in the design.
The second part is the \emph{variational} part which consists of CNOT gates used to entangle quantum states from each qubit and $R(\alpha,\beta,\gamma)$ representing the general single qubit unitary gate with three parameters $\alpha_i$, $\beta_i$ and $\gamma_i$ to be learned. 
These circuit parameters can be regarded the {weights} in the classical neural networks.
The final part is the \emph{measurement} part which will output the Pauli-$Z$ expectation values via multiple run of the quantum circuit. The retrieved values (logits) will go through classical processing such as softmax to generate the \emph{probability} of each possible class. 
The quantum measurement would be performed on first $k$ qubits where $k$ is the number of classes. 

\section{\label{sec:HybridTNVQCArchitecture}Hybrid TN-VQC Architecture}
%

\begin{figure}[tb]
\center
\includegraphics[width=.7\linewidth]{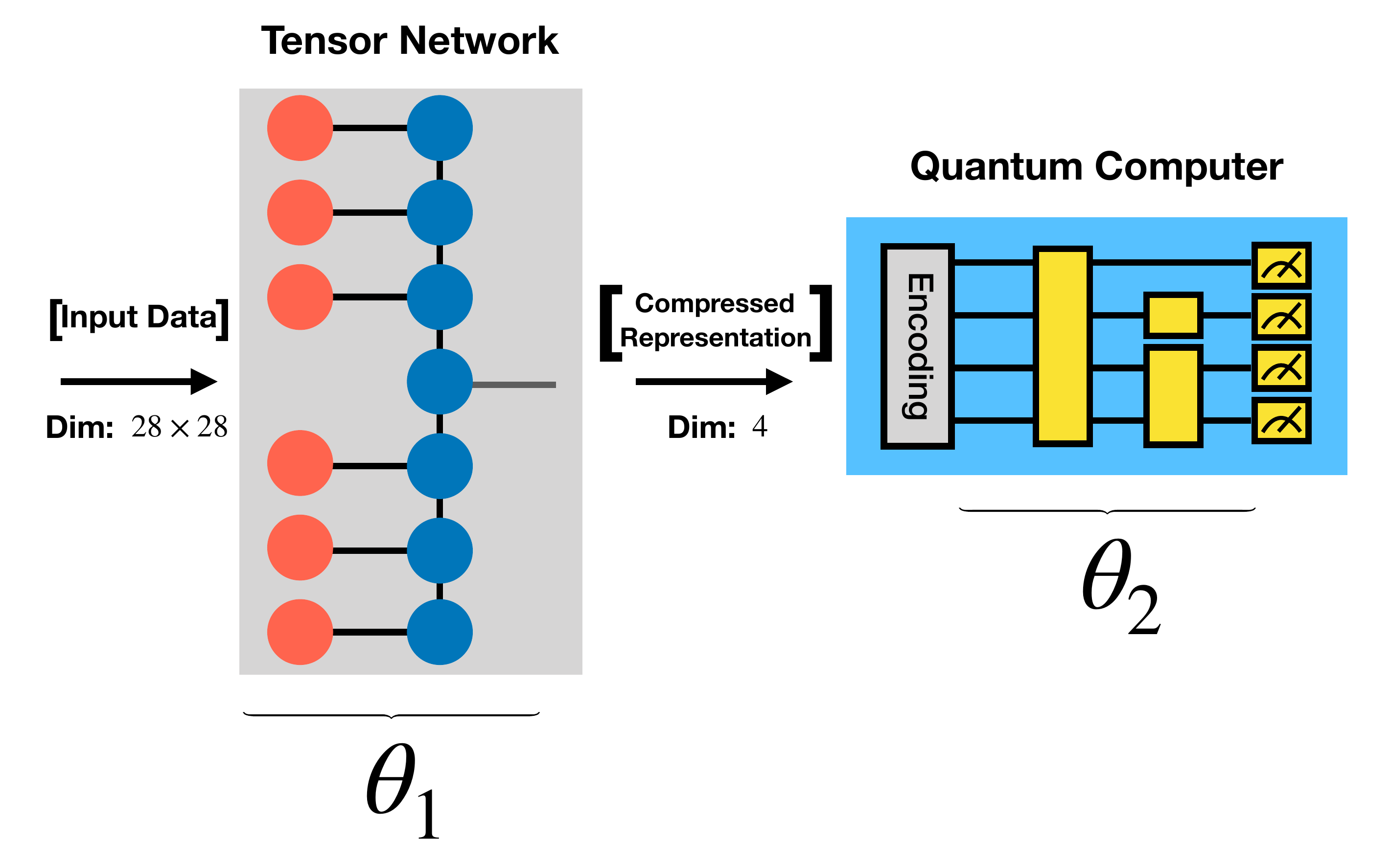}
\caption[Hybrid TN-VQC Architecture]{{\bfseries Hybrid TN-VQC Architecture.}  The tunable parameters labeled with $\theta_1$ and $\theta_2$ are the parameters for the MPS and VQC respectively. See text for details.
 }
\label{fig:Hybrid_TN_VQC_Architecture}
\end{figure}

%
%
Figure~{\ref{fig:Hybrid_TN_VQC_Architecture}} shows  the architecture of the hybrid TN-VQC model. 
The input image of $N=28\times 28=784$ pixels from MNIST or Fashion-MNIST is flatten into  a 784-dimensional vector $\mathbf{x}=\left(x_{1}, x_{2}, \ldots, x_{N}\right)$,  and  each component is normalized such that $x_i\in[0,1]$. 
The vector is mapped to a product state using the feature map~\cite{Stoudenmire:2016ve}
\begin{equation}
\mathbf{x} \rightarrow|\Phi(\mathbf{x})\rangle=
\left[\begin{array}{c}
x_{1} \\
1- x_{1}
\end{array}\right] \otimes
\left[\begin{array}{c}
x_{2} \\
1- x_{2}
\end{array}\right] \otimes \cdots \otimes
\left[\begin{array}{c}
x_{N} \\
1- x_{N}
\end{array}\right],  
\end{equation}
and further process by the MPS to generate a compressed representation. 
The feature vector is then encoded into the quantum circuit using the {variational encoding} (See \ref{encoding}).
At the end of the VQC, the quantum measurement would be performed to generate the logits  for classification.
Both the TN and VQC have tunable parameters, labeled as $\theta_1$ and $\theta_2$  respectively  in Fig.~\ref{fig:Hybrid_TN_VQC_Architecture}, which  are optimized via gradient descend methods.
Gradients of the quantum circuit parameters are calculated using the \emph{parameter-shift} method (See {\ref{sec:quantumGradient}}), which avoids  the use of finite difference calculation.
This method is similar to the computation of gradients in neural networks; therefore, the end-to-end training of this TN-VQC model follows the standard backpropagation method as in the training of deep neural networks, and no pre-trained classical model is needed. 

Since the classical and quantum parts of the model can be trained simultaneously, it allows for more flexibility in terms of  implementation on the quantum hardware. 
When more quibits are available, one simply increases the dimension of the feature vector out of the MPS to match the input of the VQC and retrain the model. 
On the other hand, the modular architecture also has the advantage that the classical and quantum parts can be reused independently.
For example, it is possible to perform transfer learning by freezing the parameters in the MPS/VQC  and training the other part to tackle different types of problems. 

\section{\label{sec:ExpAndResults}Experiments and Results}
We study the capabilities of the hybrid TN-VQC architecture by performing classification tasks on the standard benchmark dataset MNIST~\cite{lecun1998mnist} and Fashion-MNIST~\cite{xiao2017online}.
%
%
Results for the  binary classification of MNIST have been presented in Ref.~\cite{Chen:2020zv}.
Here, we perform  ternary classification   for MNIST and both  binary and ternary  classifications for Fashion-MNIST.
As a baseline, we  perform the same tasks on a  hybrid PCA-VQC model, where the PCA part serves as the simple feature extractor and the VQC as the classifier.
As a comparison, we  also present results using the MPS as a classifier to demonstrate the role of VQC in the workload. 
The computational tools we use for the simulation of variational quantum circuits and tensor networks are PyTorch~\cite{paszke2019pytorch}, PennyLane~\cite{bergholm2018pennylane} and Qulacs~\cite{suzuki2020qulacs}.
Details of the simulations such as the hyperparameters and optimizers for each experiment are summarized in \ref{details}.
\subsection{Binary Classification}
\begin{figure}[tbp]
\centering
\includegraphics[width=0.8\linewidth]{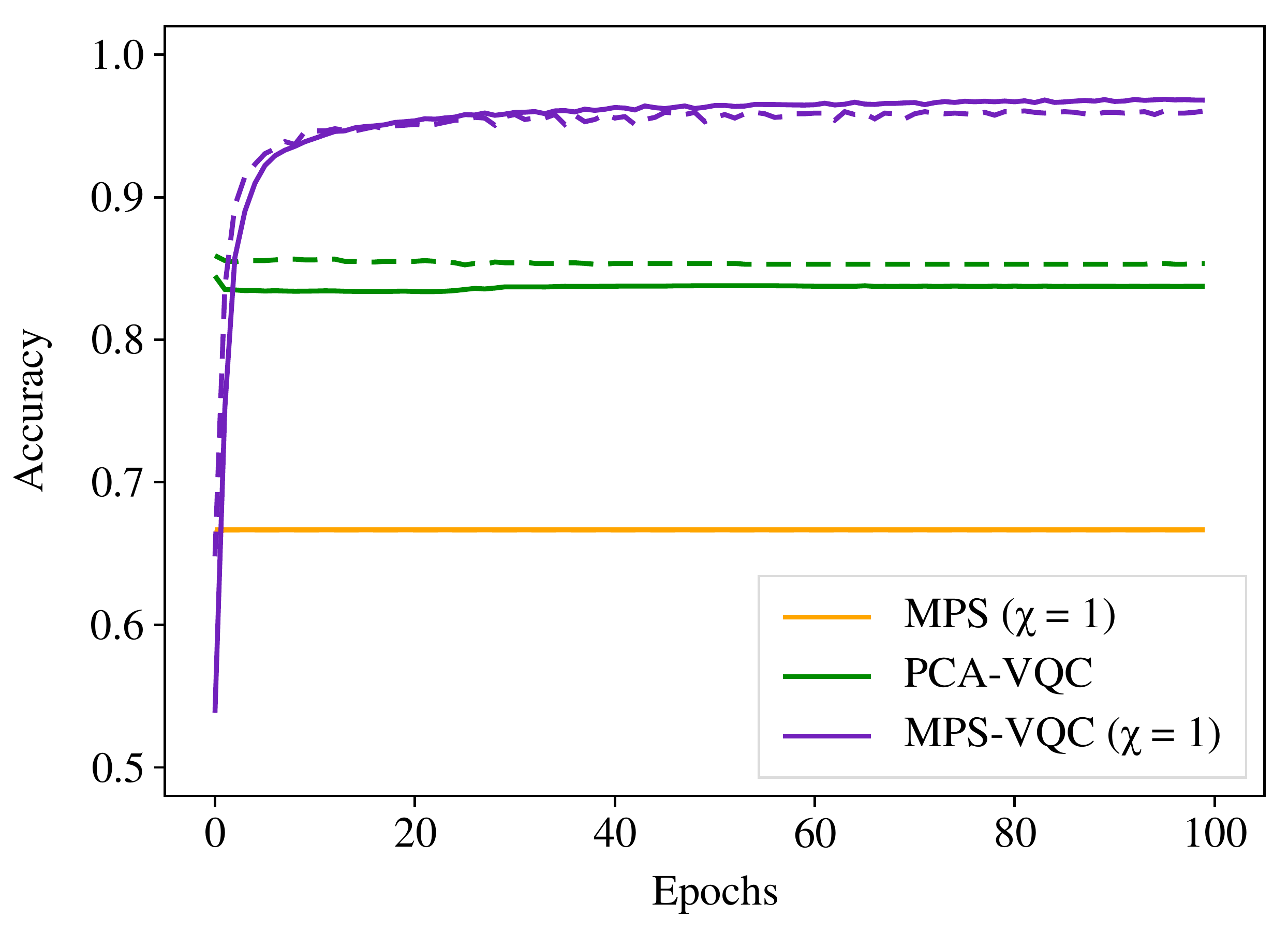}
\caption[ Binary classification of class 5 and 7.]{{\bfseries  Binary classification of class 5 and 7.}
Results of the binary classification of Fashion-MNIST (class 5 vs 7). Solid line: training accuracy. Dashed line: test accuracy}
\label{fmnist57}
\end{figure}
We perform binary classification of  the Fashion-MNIST dataset (class 5 vs 7), which is a more difficult task than the binary classification of MNIST performed in Ref.~\cite{Chen:2020zv}. 
The results from different models are shown  in Fig.~{\ref{fmnist57}}. 
For the MPS classifier and the MPS-VQC model, bond dimension $\chi=1$ is demonstrated. 
It is evident from both the training and test accuracy that such a bond dimension, while insufficient for the MPS classifier to yield good results, is  enough for the MPS-VQC hybrid model to learn properly and reach a test accuracy around $96\%$.
As the number of parameters of the VQC part is far fewer than that of the MPS, this suggests that our VQC possess greater power in classification and dominates the workload. 
It is also clear that MPS, compared to PCA, serves as a better feature extractor for a VQC discriminator. 
For a more difficult dataset, we expect that a higher bond dimension will be needed.
\subsection{Ternary Classification}
\begin{figure}[tbp]
\centering
\includegraphics[width=1.0\linewidth]{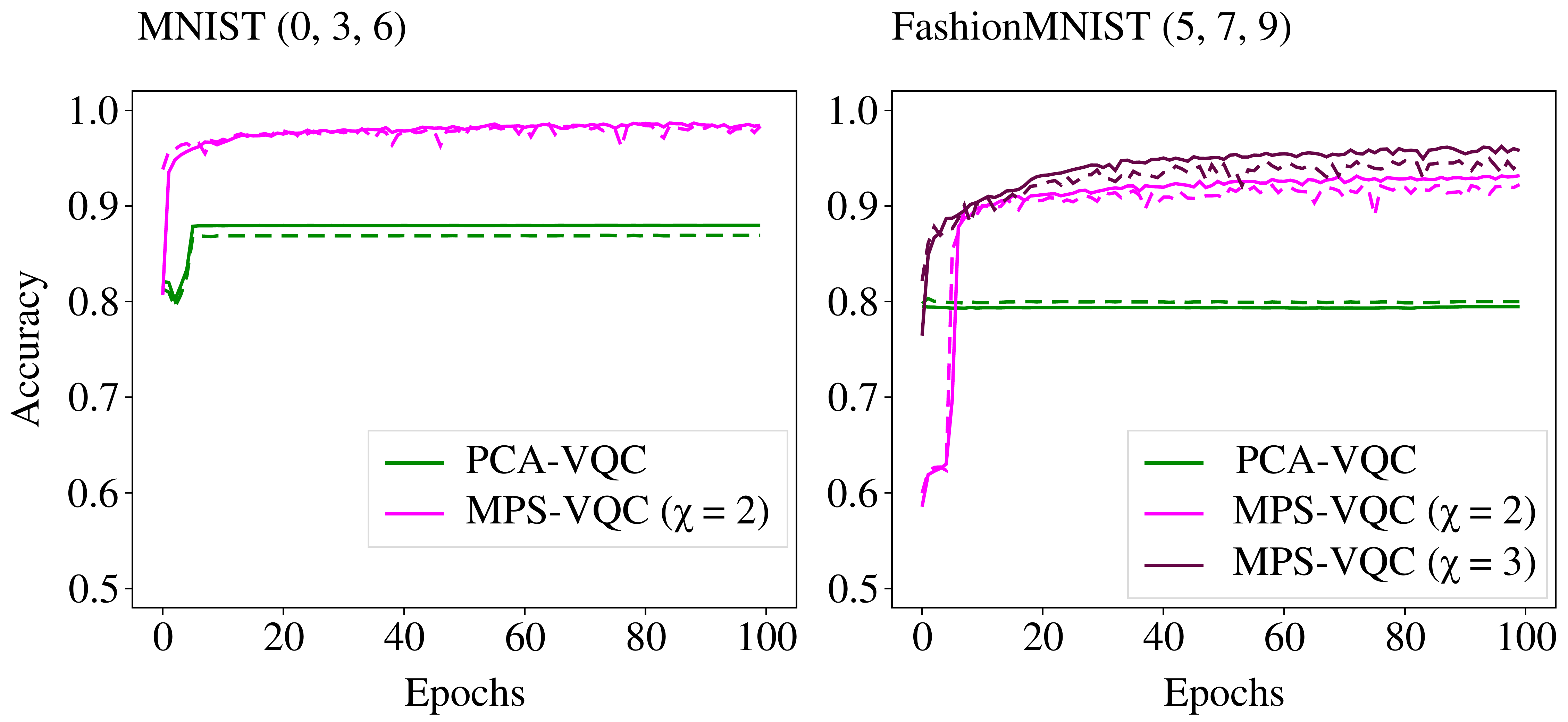}
\caption[ Ternary classification of MNIST and Fashion-MNIST.]{{\bfseries Ternary classification of MNIST and Fashion-MNIST.}
Results of the ternary classification of MNIST (class 0, 3, 6) and Fashion-MNIST (class 5, 7, 9). Solid line: training accuracy. Dashed line: test accuracy.}
\label{ternary}
\end{figure}
In the ternary classification, we consider both the  MNIST (class 0, 3, 6) and Fashion-MNIST (class 5, 7, 9) datasets.
 The results for the MPS-VQC model and the baseline model PCA-VQC are shown in~\figureautorefname{\ref{ternary}}. 
 Since ternary classification is a more difficult task, a larger  bond dimension is required  for the MPS part. 
 In terms of performance, all show that the MPS is  superior to the  PCA as a feature extractor. 
 With $\chi=2$, MPS-VQC is able to reach a test accuracy over $98\%$ in MNIST and $92\%$ in Fashion-MNIST. 
Furthermore, the representation power of an MPS feature extractor is tunable via $\chi$, which is an advantage absent in PCA. 
In the case of Fashion-MNIST, we observe a $2\%$ increase in test accuracy as the bond dimension increases,  indicating  a better data compression capability due to the increased representation power of the MPS. 
It is clear that for more complex classification problems, the performance of PCA-VQC should further deteriorate while that of MPS-VQC can be increased by increasing $\chi$.
%

%
%
\section{\label{sec:Conclusion}Conclusion}
In this work, we present a hybrid quantum-classical classifier by integrating a quantum-inspired tensor network and a variational quantum circuit. 
Such a hybrid TN-VQC architecture enables researchers to build QML applications capable of dealing with larger dimensional inputs and potentially to implement these QML models on NISQ devices with limited number of qubits and shallow circuit depth.
We further demonstrate the superiority of this framework by comparing it with the baseline study of a PCA-VQC model on ternary classification task on the MNIST and Fashion-MNIST dataset as well as a binary classification task of the Fashion-MNIST dataset.
One clear advantage is that the representation power of the trainable MPS feature extractor is tunable with the bond dimension of the tensors. 
Our results point to future  application of the hybrid TN-VQC model in different quantum machine learning scenarios and potentially implementation on NISQ devices.
%
%
Extension of this architecture to more complicated datasets such as CIFAR-10 should further test the robustness and capability of the model.
We note this requires more computing resources and better optimized simulators.
%
The proposed hybrid model can be integrated with various kinds of TNs or VQCs, given the suitable encoding methods.
For example, one can replace the MPS by other TNs with other entanglement structure such as TTN, MERA  and PEPS, whose potential in the supervised learning context has been demonstrated~\cite{Stoudenmire:2018wk,Reyes:2020fd,Glasser:2019yt}. 
They may serve also good feature extractors for datasets that contain special structure and correlations.
Another way to build a more sophisticated feature extractor is to stack multiple TN layers.
It has been shown that stacking more layers in classical deep neural networks can increase the model performance \cite{Szegedy2014GoingConvolutions, Simonyan2014VeryRecognition}. 
%
%
Attempts to build a deep convolutional tensor network~\cite{blagoveschensky2020deep} does not show the same improvement generally observed in classical deep neural network. 
 %
%
We can also replace the simple VQC in our architecture with novel VQC architectures such as  the quantum convolutional neural networks~\cite{chen2020qcnn, cong2019quantum, li2020quantum, oh2020tutorial, kerenidis2019quantum, henderson2020quanvolutional, liu2019hybrid, pesah2020absence}. 
How to apply these ideas to improve the performance of the current model is worth further investigation.

This work is supported (in part) by the U.S. DOE under grant No. DE-SC-0012704 and the BNL LDRD No.20-024 and  Ministry of Science and Technology (MOST) of Taiwan under grants No. 108-2112-M-002-020-MY3 and No. 107-2112-M-002-016-MY3.

 \appendix
 \section{\label{encoding} Encoding into Quantum States}
%
In our hybrid framework, the outputs from the classical parts need to be encoded such that they can used  by the quantum circuit.
A general $N$-qubit quantum state can be represented as:
\begin{equation}
\label{eqn:quantum_state_vec}
    \ket{\psi} = \sum_{(q_1,q_2,...,q_N) \in \{ 0,1\}^N}^{} c_{q_1,...,q_N}\ket{q_1} \otimes \ket{q_2} \otimes \ket{q_3} \otimes ... \otimes \ket{q_N},
\end{equation}
where $ c_{q_1,...,q_N}$ are complex numbers. 
They are \emph{amplitudes} of each quantum state and $q_i \in \{0,1\}$. 
The square of the amplitude $c_{q_1,...,q_N}$ represents the \emph{probability} of measurement results in  $\ket{q_1} \otimes \ket{q_2} \otimes \ket{q_3} \otimes ... \otimes \ket{q_N}$, and the total probability should sum to $1$, i.e.,
\begin{equation} 
\label{eqn:quantum_state_vec_normalization_condition}
\sum_{(q_1,q_2,...,q_N) \in \{ 0,1\}^N}^{} ||c_{q_1,...,q_N}||^2 = 1. 
\end{equation}
%
%
In this work, we choose the \emph{variational encoding} method to encode our classical data into the quantum states. 
The initial quantum state $\ket{0} \otimes \cdots \otimes \ket{0}$ first undergoes the $H \otimes \cdots \otimes H$ operation to create the unbiased state $\ket{+} \otimes \cdots \otimes \ket{+}$, where $H$ is the Hadamard gate.
 Consider a $n$-qubit system, the corresponding unbiased initial state is,
\begin{eqnarray}
\label{eqn:unbiasedInit}
    \left( H\ket{0}\right)^{\otimes n} & =\underbrace{H\ket{0} \otimes \cdots \otimes H\ket{0}}_{n}\nonumber\\
    & = \underbrace{\ket{+} \otimes \cdots \otimes \ket{+}}_{n\nonumber}\\
    & = \underbrace{\left[\frac{1}{\sqrt{2}} \left(\ket{0} + \ket{1}\right)\right] \otimes \cdots \otimes \left[\frac{1}{\sqrt{2}} \left(\ket{0} + \ket{1}\right)\right]}_{n}\nonumber\\
    & = \left[\frac{1}{\sqrt{2}} \left(\ket{0} + \ket{1}\right)\right]^{\otimes n} \nonumber\\
    & = \frac{1}{\sqrt{2^n}} \left(\ket{0} + \ket{1}\right)^{\otimes n}\nonumber \\
    & = \frac{1}{\sqrt{2^n}} \left( \ket{0}\otimes \dots \otimes\ket{0} + \dots + \ket{1}\otimes \dots \otimes\ket{1} \right)\nonumber\\
    & = \frac{1}{\sqrt{2^n}}\sum_{k= 0}^{2^n - 1}\ket{k},
\end{eqnarray}
%
%
This initial quantum state will first go through the {encoding} part which consists of $R_y$ and $R_z$ rotations. 
These rotation operations are parameterized by the input vector $\vec{x} = \left( x_1, x_2, \cdots x_n \right)$. 
 On the $i$-the  qubit with $i = 1 \cdots n $, $R_y$ rotates the state by an angle of  $\arctan(x_i)$ and  $R_z$ by $\arctan(x_i^2)$.
The encoded state is then processed with the {variational} quantum circuits with optimizable parameters, as shown in the dashed-box in Fig.~{\ref{Fig:Basic_VQC_Hadamard}}. 
%
\section{\label{sec:quantumGradient}Calculation of Gradients of Quantum Functions}
%
Here the models are trained via gradient-descent methods widely used in training the deep neural network. 
To calculate the gradients with respect to the parameters of quantum circuits, we employ the \emph{parameter-shift} method \cite{schuld2019evaluating, bergholm2018pennylane, mitarai2018quantum}.
Given the knowledge of computing the expectation values of an observable $\hat{P}$ on quantum function,
\begin{equation}
f\left(\mathbf{x} ; \theta_{i}\right)=\left\langle 0\left|U_{0}^{\dagger}(\mathbf{x}) U_{i}^{\dagger}\left(\theta_{i}\right) \hat{P} U_{i}\left(\theta_{i}\right) U_{0}(\mathbf{x})\right| 0\right\rangle=\left\langle x\left|U_{i}^{\dagger}\left(\theta_{i}\right) \hat{P} U_{i}\left(\theta_{i}\right)\right| x\right\rangle,
\end{equation}
where $\mathbf{x}$ is the classical input vector (e.g. the output values from the PCA or MPS parts), $U_0(x)$ is the quantum encoding routine to prepare the classical value $x$ into a quantum state, $i$ is the circuit parameter index for which the gradient is to be evaluated,
and $U_i(\theta_i)$ represents the single-qubit rotation generated by the Pauli operators $X, Y , Z$. It can be shown \cite{mitarai2018quantum} that the gradient of this quantum function $f$ with respect to the parameter $\theta_i$ is
\begin{equation}
    \nabla_{\theta_i} f(x;\theta_i) = \frac{1}{2}\left[ f\left(x;\theta_i + \frac{\pi}{2}\right) - f\left(x;\theta_i - \frac{\pi}{2}\right)\right].
    \label{eq:quantum gradient}
\end{equation}
%
%

\section{\label{details}Computational details}
Here we summarize the optimizer and hyperparameters in our experiments. 
\begin{itemize}
\item MPS classifier:
In experiments where MPS alone is used as a classifier, the optimizer is Adam~\cite{kingma2014adam} with a learning rate of $ 0.001$ and batch size of $100$.
\item PCA-VQC model:
In our baseline model, PCA is used to reduce the input dimension of $28 \times 28 = 784$ into a four-dimensional vector and is implemented with the Python package scikit-learn \cite{scikit-learn}.
For the VQC classifier, the optimizer is RMSProp \cite{Tieleman2012} with the hyperparameters: learning rate $ = 0.01$, $\alpha = 0.99$ and $\epsilon = 10^{-8}$.
\item MPS-VQC hybrid model:
In the hybrid TN-VQC architecture, the optimizer is Adam~\cite{kingma2014adam} with a learning rate of $ 0.0001$ and batch size of $50$. 
The VQC  consists of four  variational blocks (dashed-line box in \figureautorefname{\ref{Fig:Basic_VQC_Hadamard}}). 
The number of quantum circuit parameters is $4 \times 4 \times 3 = 48$ in this setting. 
The quantum measurement at the final part of a VQC depends on the number of classes to be classified. 
For binary classification tasks, the first two qubits are measured, whereas for ternary classification, we perform  quantum measurement on the first three qubits. 
\end{itemize}
In all cases, we use cross entropy loss as our loss function. 
\section*{References}
\bibliographystyle{iopart-num}
\bibliography{bib/qecc,bib/nisq,bib/vqc,bib/qml_examples,bib/qml_general,bib/machinelearning,bib/tool,bib/TN,bib/tn_ml,bib/qc}

\end{document}